\documentclass[aps,prb,twocolumn,superscriptaddress,showpacs]{revtex4}
\usepackage[dvipdfm]{graphicx}

\begin{document}

\title{Phase diagram for the one-dimensional Hubbard-Holstein model:
A density-matrix renormalization group study}

\author{Masaki Tezuka}
\email{tezuka@cms.phys.s.u-tokyo.ac.jp}
\thanks{Present address:
Department of Physics, Tokyo Institute of Technology,
Ookayama, Tokyo 152-8551, Japan}
\affiliation{Department of Physics, University of Tokyo,
Hongo, Tokyo 113-0033, Japan}
\author{Ryotaro Arita}
\affiliation{RIKEN, Wako, Saitama 351-0198, Japan}
\author{Hideo Aoki}
\affiliation{Department of Physics, University of Tokyo,
Hongo, Tokyo 113-0033, Japan}

\date{\today}

\begin{abstract}
Phase diagram of the Hubbard-Holstein model in the coexistence 
of electron-electron and electron-phonon interactions has been 
theoretically obtained with the density-matrix renormalization 
group method for one-dimensional (1D) systems, where 
an improved warm-up (the recursive sweep) procedure 
has enabled us to calculate various correlation functions.  
We have examined the cases of (i) the systems half-filled by electrons 
for the full parameter space spanned by the electron-electron 
and electron-phonon coupling constants and the phonon frequency, 
(ii) non-half-filled system, and (iii) trestle lattice.  
For (i), we have detected 
a region where both the charge and on-site pairing correlations 
decay with power-laws in real space, which suggests a metallic behavior. 
While pairing correlations are not dominant in (i), we have 
found that they become dominant as the system is doped in (ii), 
or as the electronic 
band structure is modified (with a broken electron-hole symmetry) 
in (iii) in certain parameter regions.  
\end{abstract}


\maketitle

\section{Introduction}
In condensed-matter physics, the electron-phonon interaction and 
electron-electron interaction are two of the fundamental 
interactions.  While the problem is hard enough even when 
only one of them exists, it is a most challenging 
problem to consider what happens when both of the interactions coexist.  

A central problem is the ground-state phase diagram --- how 
various phases arise in the coexistence of two kinds of 
interactions that are quite different in nature.  
Of particular interest is superconductivity.  
Strong electron-electron repulsive interactions introduce spin fluctuations 
in the electronic system, which can mediate pairing interactions 
for the electrons to make them 
condense into superconducting states with anisotropic Cooper pairs.\cite{1987PhRvL..58.2794E}
The superconducting phase has to, however, compete 
with density-wave phases such as the spin-density wave (SDW), 
which also come from the electron-electron interaction.\cite{1988PhRvL..60..944S}
On the other hand, 
the electron-phonon interaction mediates an attraction
between electrons, which can make the electrons condense into superconducting 
states with isotropic Cooper pairs.  
This time the superconducting phase has to compete 
with density-wave phases such as the charge-density wave (CDW) 
arising from the electron-phonon interaction.
The electron-phonon interaction is the coupling of the conduction
electrons to the lattice structure, and it can also work to make
the system undergo a Peierls transition, where the
lattice is deformed so that the electrons are not conducting. 

So it is quite a nontrivial problem to consider the ground-state 
phases when the two interactions coexist and are both strong.  
A prototypical model representing such a situation 
is the Hubbard-Holstein model, where the Hubbard model for the 
electron correlation is coupled to (Einstein) phonons. 
The model is characterized by three physical parameters: 
(a) the on-site electron-electron 
repulsion $U$, 
(b) the phonon frequency $\omega_0$, 
and (c) the electron-phonon coupling $\lambda$, 
where the unit of energy is the electronic transfer $t\propto$ 
electronic bandwidth. 
Of particular interest is the intermediate regimes: 
(a) the regime ($\hbar \omega_0 \sim t$) 
intermediate between the adiabatic 
($\hbar \omega_0 \ll W$) and anti-adiabatic ($\hbar \omega_0 \gg W$) limits, 
(b) the regime ($U\sim \lambda$) where the electron-electron Coulomb repulsion 
and the phonon-mediated attraction are similar in magnitude.  
The problem has in fact been studied in various approaches.
\cite{
PhysRevB.10.1896, 1977PhRvB..16.3943H, PhysRevB.27.4302, guinea1983lmb,
PhysRevB.29.4230, PhysRevB.31.6022, 1986PhRvB..34.7429V, nasu1987mpt,
1988PhRvB..3710068V, PhysRevLett.60.2089, PhysRevLett.64.323, takada1996shf, 
PhysRevB.54.2410, 1997PhRvB..5514886L, 1998PhRvB..57.5051P,
1998PhyD..113..307P, 1999PhRvB..59.1444Y, 2001PhRvB..64i4507B,
takada2003pmp, 2004PhRvB..69p5115F, 2004PhRvB..69x5111M,
2005PhyB..359..795K, 2005PhRvL..95i6401C, 2007PhRvB..75p1103T}

This is by no means a theoretical curiosity, since 
there are various classes of materials in which 
both the electron-electron and electron-phonon interactions 
are simultaneously strong.  
A typical example is the solid fullerene doped with alkali-metal 
atoms, ($A_3\mbox{C}_{60}$, $A=\mbox{K}, \mbox{Rb}, ...$),
\cite{1991Natur.350..600H,1991Natur.352..222T,1997RvMP...69..575G}
where the $\mbox{C}_{60}$ fullerene molecules are aligned in a
face-centered cubic (fcc) lattice, for which 
the alkali atoms supply electrons making the conduction band
half-filled.  
In this material the intramolecular phonon modes are known to have 
high frequencies ($\sim 0.2$ eV), which 
couple strongly to the conduction electrons.  
The material is a superconductor whose transition temperature 
$T_C$ is the highest to date among the carbon compounds.
It is an interesting problem to consider 
how the half-filled electron band becomes metallic,
rather than becoming CDW or SDW insulators from 
coexisting electron-electron and electron-phonon interactions, 
and exhibits superconductivity.  
The electronic bandwidth of $A_3\mbox{C}_{60}$ can be 
controlled by e.g. changing the alkali metal species $A$, 
where $T_C$ is observed to increase with the lattice constant.  
Organic chemists have also fabricated fullerene derivatives in the 
shape of a shuttlecock, where the molecules 
stack to form an array of 1D chains,\cite{sawamura2002scm}
where the separation between $\mbox{C}_{60}$ molecules in a chain 
is controllable with the choice of functional groups
attached to $\mbox{C}_{60}$, and the system is shown to have 
stronger electron correlation.\cite{okada2004ess}

Given these backgrounds, 
the purpose of the present paper is to obtain the ground-state phase 
diagram when the electron-phonon and electron-electron 
interactions coexist over the whole 
parameter space 
spanned by $U, \lambda, \hbar\omega$ that includes the 
intermediate regimes ($\hbar \omega \sim W$; $U\sim \lambda$).  
We adopt the Hubbard-Holstein model, one of the
simplest models for the coexisting 
electron-electron and electron-phonon interactions.    
We want to treat these interactions on an equal footing, 
especially in the intermediate regimes.  
Phonons mediate attraction between electrons, 
but here we have adopted the dispersionless,
Einstein phonons that do not directly propagate along the chain.
While these phonons cannot directly mediate the pairing force between
electrons across different sites, we can still expect the phonons to 
contribute to the pairing, because electrons
hop between neighboring sites. 
So the phonon self-energy $\Sigma(q,\omega)$ should 
depend not only on $\omega$ but also on $q$.  
Specifically, phonons with $q\sim \pi/a$ 
($a$: the lattice constant) as well as phonons with $q\sim 0$ 
should be strongly renormalized by the coupling to the electrons, 
because the Fermi points reside at $k=\pm \pi/2a$.
This is another reason why the electron and phonon degrees
of freedom should be treated on an equal footing.
 
So a special care has to be taken in choosing the method, 
since the method has to (a) take account of the phonons 
without assuming the adiabatic or anti-adiabatic limits, 
and (b) take account of the electrons where charge gaps 
around the half-filled electronic band 
(a region of interest) can be described.  
We have adopted the density-matrix renormalization group
(DMRG) as an appropriate method.  

DMRG was originally developed for spin systems
\cite{White1992PhRvL..69.2863W,White1993PhRvB..4810345W}
and applications to electron systems subsequently followed.
\cite{1994PhRvL..73..882N,1995PhRvL..75..926H}
The basic idea is the following: 
We iteratively add sites to enlarge a one-dimensional system,
which is approximately represented as two connected \textit{subchains}
with truncated Hilbert spaces optimized for a target state of the whole system.
In adding a site to a subchain, we calculate the partial density matrix
for the target state wave function, diagonalize it, and
retain eigenstates that correspond to a limited number of
largest eigenvalues, to represent the new Hilbert space for the new subchain.
While the DMRG was developed for electron systems, 
Jeckelmann and White extended the formalism to 
incorporate phonons in DMRG.\cite{Jeckelmann1998PhRvB..57.6376J} 
We have here adopted this formalism, with an improved 
algorithm. 

In constructing a phase diagram, a most direct method is to 
actually calculate various correlation functions.  
However, the inclusion of phonons, which are bosons, makes 
the calculation enormously heavy, and we have to 
somehow overcome this difficulty.   
Here we have adopted 
an improved warm-up (the recursive sweep) procedure 
for the DMRG developed by one of the 
present authors,\cite{2007JPSJ...76e3001T} which has enabled us to actually calculate 
correlation functions (including the pairing correlation) 
for the first time for the Hubbard-Holstein model.  
While taking 1D systems is a theoretical device (apart from 
the above mentioned quasi-1D fullerides), 
it is theoretically intriguing to look at functional 
forms of correlation functions in the 
coexistence of 
electron-electron and electron-phonon interactions.  
The Mermin-Wagner theorem dictates that
continuous symmetries of the Hamiltonian are not broken 
in the ground states of infinite 1D systems.  
Hence the correlation functions for the phases 
with broken continuous symmetry should decay with distance.  
We can still compare the exponents of the decay, as 
in the Tomonaga-Luttinger theory for purely electronic systems, 
to determine the dominant correlation, which may become
long-ranged once we go to higher dimensions.

Now, there are many existing theoretical studies on the 
Hubbard-Holstein model,
but the nature of the model has yet to be fully established.  
There have been competing results even on the ground-state phase diagram
at half-filling.  Particular issues are: 
\begin{enumerate}
\item For large enough $U$ we can expect the ground state to be 
an SDW, while for large enough $\lambda$ we can expect the ground state to be 
a CDW.  An interesting problem is whether there exists 
a metallic phase between the two density-wave phases.  
\item Whether a na\"{\i}ve expectation is valid, 
where one imagines that, 
for the coexisting electron-electron and electron-phonon interactions, 
the net interaction will simply be the difference between them.  
This problem becomes especially interesting in the intermediate regime.  
\item Whether the metallic phase, if any, 
can possibily become superconducting.  
For pairing symmetries, we have here considered all the 
possibilities of
\begin{itemize}
\item sSC: on-site spin-singlet pairing,
\item pSC: nearest-neighbor spin-triplet pairing,
and
\item dSC: nearest-neighbor spin-singlet pairing.
\end{itemize}
It may sound peculiar when we say $p$- or $d$-wave
superconductivity for 1D systems,
but a nearest-neighbor spin-triplet (singlet) pairing
corresponds to a $p$($d$)-wave pairing in two or higher dimensions, 
so we adopt this nomenclature.
\item How the phase diagram depends on the 
electron band filling, or the 
lattice structure (i.e., the electronic band structure) of quasi-1D systems.
\end{enumerate}

In order to resolve these issues 
we have obtained the correlation functions for long chains with DMRG 
for the cases of (i) the systems half-filled by electrons 
for the full parameter space spanned by the electron-electron 
and electron-phonon coupling constants and the phonon frequency, 
(ii) non-half-filled system, and (iii) trestle lattice.  
For (i), we have detected 
a region (where the electron-phonon interaction is stronger than 
$U$ but not too strong) 
where both the charge and on-site (sSC) pairing correlations 
decay arithmetically in real space, which suggests a metallic behavior. 
While pairing correlations are not dominant in (i), we have 
found that they become dominant as the system is doped in (ii), 
or as the electronic 
band structure is modified (with a broken electron-hole symmetry) 
in (iii) in certain parameter regions.  

Organization of the paper is as follows: 
In section II we introduce the Hubbard-Holstein model, and
briefly review the open questions about the phase diagram.
In section III we explain the numerical method for calculating
correlation functions.
Results are presented in sections IV (half-filled chain), 
V (doped chain) and VI (undoped trestle lattice). 
Concluding remarks are given in section VII.

\section{Model}
\begin{figure}
\begin{center}
\includegraphics[width=8.6cm]{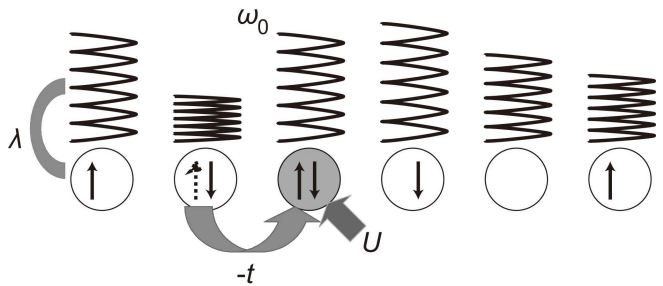}
\caption{The Hubbard-Holstein model.
Each atom consists of an electron site (circle) and a local phonon 
with frequency $\omega_0$ (spring).
Arrows denote up and down spin electrons, which are 
coupled to the phonons with a strength $\lambda$.
An electron can hop between neighboring sites with the
hopping amplitude $t$, and feels an 
on-site electron-electron repulsion when encountered 
with another electron of opposite spin on the same site.
}
\label{fig:HHmodel}
\end{center}
\end{figure}

We take the Hubbard-Holstein model, where 
the electronic part of the Hamiltonian is the Hubbard model 
that in itself exhibits various phases
according to the values of the electron band filling $n$ and the 
short-range (on-site) electron-electron interaction $U$, 
dimensionality, and the lattice structure.  
Phonons are introduced as a local harmonic oscillator for 
each site, to which 
the electrons are coupled as shown schematically in Fig. \ref{fig:HHmodel}.  
Thus the phonons are dispersionless (i.e., Einstein phonons).  
Inclusion of on-site phonons adds two parameters to the Hubbard model:
the electron-phonon coupling $g$ and
the phonon frequency $\omega_0$.
We can characterize the strength of the electron-phonon coupling, 
\[
 \lambda \equiv 2g^2/\hbar\omega_0,
\]
which is the attraction 
mediated by phonons between two electrons on the same site 
in the anti-adiabatic ($\omega_0/t\rightarrow\infty$) limit.  
We take this as the measure of the electron-phonon coupling to 
facitilate comparison with the electron-electron repulsion $U$.  
We set $\hbar=1$ hereafter.

The Hamiltonian is given as
\begin{eqnarray}
H&=&-\sum_{i,j,\sigma}t_{ij}(c_{i\sigma}^\dag c_{j\sigma} + c_{j\sigma}^\dag c_{i\sigma})
+ \sum_i U n_{i\uparrow} n_{i\downarrow}
\nonumber\\
&+& \sum_{i,\sigma} g n_{i\sigma}(\tilde a_i+\tilde a_i^\dag)
+ \sum_i \hbar\omega_0 \tilde a_i^\dag \tilde a_i.
\label{eqn:HHorig}
\end{eqnarray}
Here, 
$c_{i\sigma}$ annihilates an electron with spin $\sigma(=\uparrow,\downarrow)$ at site $i$, 
$n_{i\sigma}=c_{i\sigma}^\dag c_{i\sigma}$ is the electron number, and
$\tilde a_i$ is the phonon annihilator at site $i$.
If we consider the phonon as a harmonic oscillator where a mass $M$ is
attached to a spring with a spring constant $K$ with the phonon frequency
$\omega_0=\sqrt{K/M}$,
the displacement $x_i$ is written as
\begin{equation}
x_i=\frac{\hbar}{2M\omega_0}(\tilde a_i+\tilde a_i^\dag).
\end{equation}
The electron-phonon coupling is introduced as a coupling 
of the local electron
number, $n_i\equiv\sum_\sigma n_{i\sigma}$, 
with the lattice displacement $x_i$.

The electron band filling is definded as $\overline n_\sigma = n/2$, 
and, by introducing $a\equiv \tilde a-g\overline n_\sigma/\hbar\omega_0$,
we can rewrite (\ref{eqn:HHorig}), up to a constant, as
\begin{eqnarray}
H&=&-\sum_{i,j,\sigma}t_{ij}(c_{i\sigma}^\dag c_{j\sigma} + c_{j\sigma}^\dag c_{i\sigma})\nonumber\\
&+& \sum_i U (n_{i\uparrow}- \overline n_\uparrow) (n_{i\downarrow}-\overline n_\downarrow)
\nonumber\\
&+& \sum_{i,\sigma} g (n_{i\sigma}-\overline n_\sigma) (a_i+a_i^\dag)
+ \sum_i \hbar\omega_0 a_i^\dag a_i.
\label{eqn:HHdiff}
\end{eqnarray}
In this form, the relevant 
electron occupation is the deviation from the average value.

The Hubbard-Holstein model (\ref{eqn:HHorig}) has three independent parameters:
\[
U/t,\lambda/t,\omega_0/t
\]
as schematically displayed in Fig. \ref{fig:1dHolHubfig}.  
In this parameter space, we are interested in the regimes 
(i) away from the adiabatic ($\omega_0/t\ll 1$) or anti-adiabatic
($\omega_0/t\gg 1$) limits,
and (ii) comparable $U \sim \lambda$.  
To summarize the existing theoretical results on the
1D half-filled Hubbard-Holstein model, 
we observe the following issues:
\begin{itemize}
\item Two of the recent studies on this model,
one with the Lang-Firsov transformation\cite{takada2003pmp}
and the other with DMRG,\cite{2004PhRvB..69p5115F} 
have different conclusions as to where this model becomes metallic.
The former indicates a considerably wide metallic region 
between the CDW ($U\ll\lambda$) and SDW ($\lambda\ll U$) phases,
while the latter indicates a closing of the charge gap only at a 
quantum critical point between the CDW and SDW insulators.  
\item While the strong-coupling expansion \cite{PhysRevB.31.6022}
indicates that superconductivity cannot be dominant
for finite $\omega_0$,
a study with some ansatz for phonons \cite{guinea1983lmb}
indicates a dominant superconductivity
near the $U=0$ axis on the $U$--$\lambda$ phase diagram,
and another QMC study for the charge structure factor
suggests a phase diagram with a superconducting region 
near the $U=\lambda$ line between the CDW and SDW phases.
\end{itemize}

On the other hand, the Holstein model
corresponds to the $U=0$ plane in the phase diagram
of the Hubbard-Holstein model, 
and picks up the effect of the electron-phonon
interaction in making the electronic system insulating.
In one \cite{Jeckelmann1999PhRvB..60.7950J,1996PhRvB..54.8981S}
and two dimensions,
and in the limit of infinite dimension,\cite{1993PhRvB..48.6302F,2003PhRvL..91r6405C}
there are extensive studies on the transition between
(i) the small-polaron regime where the electrons are self-trapped by the lattice
through a stronger electron-phonon interaction, 
and
(ii) the large-polaron regime where the electrons move around 
exciting local phonons through a weak electron-phonon interaction.\cite{PhysRevLett.36.323}
When the electron-electron interaction is turned on in such an electron-phonon 
system, the interplay between the effects of the electron-phonon and
electron-electron interactions again becomes interesting, which 
has not been fully understood.

\section{Method}
As we have seen, the ground-state phase diagram for the 1D
Hubbard-Holstein model is controversial.  
The most clear-cut way for examining the competition of
various phases is to look at the correlation functions on long systems,
which has not previously been done properly.
So here we calculate correlation functions in real space.  
A care has to be taken in examining correlation functions
in 1D quantum systems:
strong quantum fluctuations repress continuous symmetry-broken
long-range orders in the ground state.\cite{PhysRevLett.17.1133,PhysRev.158.383}
However, if one type of correlation function decays more slowly than others,
we can identify the correlation to be dominant.
Exact results for purely electronic systems
on various integrable models show that
diagonal [$\langle c^\dag(r) c(r) c^\dag(0) c(0)\rangle$] and
pair [$\langle c^\dag(r) c^\dag(r) c(0) c(0)\rangle$]
correlations decay
with a power-law ($r^{-\eta}$, where $\eta$ is an exponent)
or exponentially ($e^{-r/\xi}$, where $\xi$ is the correlation length)
at large distances $r$.

In the case of charge and spin correlations,
the charge (spin) correlation function decays with a power-law
when the charge (spin) excitation is gapless, $\Delta_{\rm charge}=0$ 
($\Delta_{\rm spin}=0$),
while the correlation decays exponentially when the charge (spin) excitation has a gap
$\Delta_{\rm charge}>0$ ($\Delta_{\rm spin}>0$).
To be precise, these correlations can have logarithmic corrections or
additional terms that decay faster than the main term.
This is one reason why we need long chains 
to identify the dominant phase by numerically calculating
the decay of correlation functions.  
Here we compare the real-space behavior of various correlation functions
in the ground state for a chain having as many (typically 64) 
sites as numerical calculation is possible for 
model parameters away from adiabatic and anti-adiabatic limits.  
For these reasons, we adopt the DMRG with
the pseudo-site method
\cite{Jeckelmann1998PhRvB..57.6376J}
to calculate the
ground-state correlations.  We adopt the open-boundary condition
for a better DMRG convergence.
The recursive sweep initialization \cite{2007JPSJ...76e3001T} adds two
sites for both the electron and phonon degrees of freedom at
each infinite-algorithm DMRG step, while benefiting from the
reduced size of the Hilbert space in the pseudo-site method.
This also contributes to better convergence with less
computational resources, which has allowed us to compute
correlation functions on lattices with large number of sites.

\section{Result: Half-filled Hubbard-Holstein chain}
\subsection{Parameter space}
\begin{figure}
\begin{center}
\includegraphics[width=5.6cm]{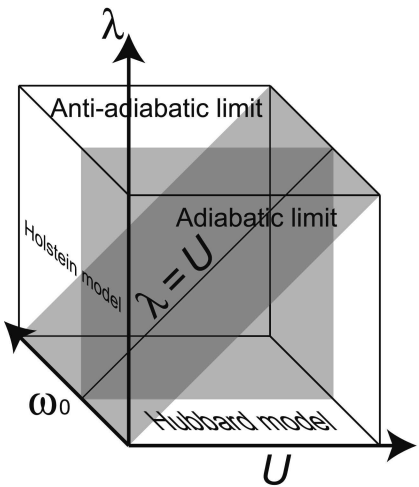}
\caption{Parameter space for 
the half-filled Hubbard-Holstein model spanned by 
the on-site electron-electron repulsion $U$, 
the phonon frequency $\omega_0$, 
and the electron-phonon coupling $\lambda$, 
where the unit of energy is the electronic transfer $t$. 
The model reduces to the Hubbard model when $\lambda=0$,
and to the Holstein model when $U=0$.
The $\omega_0/t\rightarrow0$ limit is the adiabatic limit,
while the $\omega_0/t\rightarrow\infty$ limit is the anti-adiabatic limit.
}
\label{fig:1dHolHubfig}
\end{center}
\end{figure}

First we investigate half-filled systems 
for the parameter space shown as a bird's eye view 
in Fig. \ref{fig:1dHolHubfig} 
(with $t=1$ taken as the unit of energy), 
while the doped case is treated in \S\ref{sec:doped}.   

We calculate charge, spin and pair correlation functions for 
this parameter space. 
For the pairing symmetry we consider 
the on-site, spin-singlet pair (sSC) 
(which corresponds to $s$-wave superconductivity in higher spatial dimensions), 
spin-singlet pair across the neighboring sites (dSC) 
($d$-wave in higher dimensions), 
and triplet pair across neighboring
sites (pSC) ($p$-wave in higher dimensions). 
Their operator forms are:
\begin{itemize}
\item Charge:
$\langle n_i n_j\rangle-\langle n_i\rangle\langle n_j\rangle$,
\item Spin:
$\langle S^z_i S^z_j \rangle$,
\item sSC:
$\langle \Sigma^\dag_i \Sigma_j\rangle$,
\item pSC:
$\langle \Pi^\dag_i \Pi_j\rangle$,
\item dSC:
$\langle \Delta^\dag_i \Delta_j\rangle$,
\end{itemize}
where $n_i\equiv\sum_\sigma n_{i\sigma}$,
$S^z_i\equiv(n_{i\uparrow}-n_{i\downarrow})/2$,
$\Sigma_i\equiv c_{i\uparrow}c_{i\downarrow}$,
$\Pi_i\equiv (c_{i\uparrow}c_{i+1\downarrow}+c_{i\downarrow}c_{i+1\uparrow})/\sqrt2$,
and
$\Delta_i\equiv (c_{i\uparrow}c_{i+1\downarrow}-c_{i\downarrow}c_{i+1\uparrow})/\sqrt2$.

We first display typical dependence of correlation functions
on the real-space distance along the chain.
The calculations have been done on $L=64$-site, half-filled 
Hubbard-Holstein chains,
with pseudo-site DMRG for at least $N_b=3$ phonon pseudo-sites per full site,
unless otherwise indicated.  
At least $10$ finite-system sweeps have been done with $m>400$ states retained
in more than two final iterations after the infinite-finite method warm-up.
The results do not change significantly 
when we take larger $N_b\geq 4$ or $m>500$.
The maximum discarded weight of the partial density matrix
in the final sweep is below $10^{-5}$ when $m>400$ and
typically around $10^{-7}$ when $m>500$.

\subsection{The case of larger $\omega_0/t$}
\begin{figure}
\begin{center}
\includegraphics[width=8.6cm]{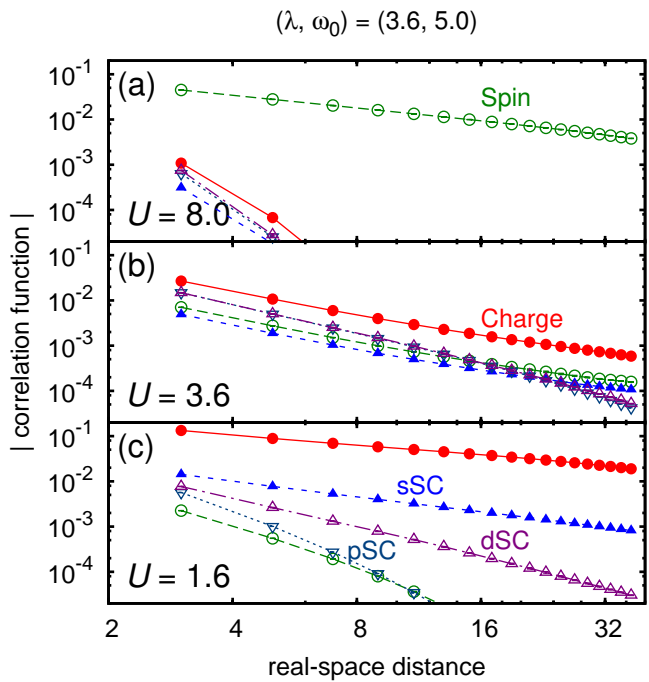}
\caption{(Color online) Log-log
plot of correlation functions against the real-space distance 
for $(\lambda, \omega_0) = (3.6, 5)$
and 
$U=$ (a) $8.0$, (b) $3.6$ and (c) $1.6$ 
for the half-filled Hubbard-Holstein model.}
\label{fig:1dcor_large}
\end{center}
\end{figure}
First we consider the region where the phonon frequency
is greater than the electron
hopping $t$.
We fix the electron-phonon coupling $\lambda$ and phonon frequency $\omega_0$
at $(\lambda, \omega_0) = (3.6, 5.0)$.
We observe in Fig. \ref{fig:1dcor_large}(a) that 
only the spin correlation decays with a power-law
for $U$ much larger than $\lambda$.  The exponent is near unity.

As $U$ is decreased to around $\lambda$,
the charge correlation begins to decay with a power-law,
as plotted in Fig. \ref{fig:1dcor_large}(b) for $U=3.6=\lambda$.
The exponent for the spin correlation is greater than in 
Fig. \ref{fig:1dcor_large} (a).
Here we note that the on-site pair correlation also decays with a power-law.

Finally, for a much smaller value of $U=2.6$ in Fig. \ref{fig:1dcor_large}(c), 
the spin correlation decays no longer with a power-law,
but almost exponentially at large distances, which becomes
more manifest as $U$ is further decreased. 
There, both CDW and on-site pair correlations still decay with power-laws.  
So we observe a power-law spin correlation when the electron-electron
interaction is stronger than the electron-phonon interaction, 
or power-law charge and on-site pair correlations when the electron-electron
interaction is weaker.

\begin{figure}
\begin{center}
\includegraphics[width=8.6cm]{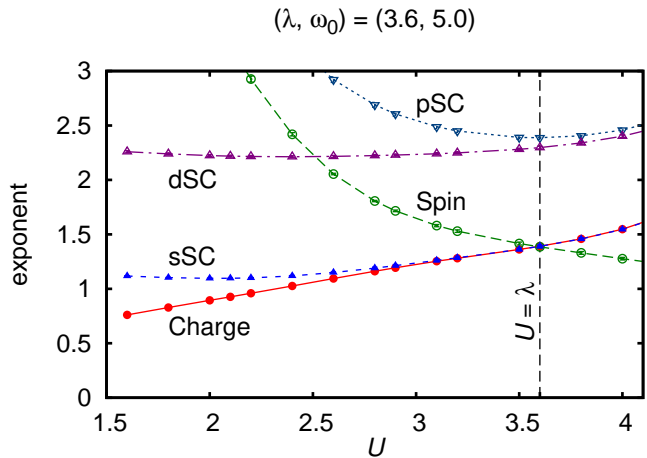}
\caption{(Color online) Exponents of correlation functions for
the Hubbard-Holstein model with $(\lambda, \omega_0)=(3.6, 5.0)$.}
\label{fig:1dexp_5.0}
\end{center}
\end{figure}

Next we compare the exponents of the correlations.
We fit the correlations as functions of the real-space distance $r$,
assuming the behavior $\langle O^\dag(x)O(x+r)\rangle\sim r^{-\eta}$,
and determine the exponent $\eta$.
The results are plotted in Fig. \ref{fig:1dexp_5.0}.
We can see that charge and on-site pair correlations have similar exponents,
but the exponent for the charge correlation is always smaller
than that for the on-site pair correlation. 
So the charge correlation is dominant when $U/t$ is smaller than
around $3.5$ where $\lambda=3.6$,
while the spin correlation is dominant when $U/t$ is larger.

For the repulsive ($U>0$) Hubbard model,
we have only the spin correlation decaying as $r^{-1}$.
Now, for the purely electronic Hubbard model, 
it has been known \cite{1972PThPh..48.2171S,1974PThPh..52.1716N} that 
we can convert the repulsive ($U>0$) Hubbard model into the 
attractive ($U<0$) Hubbard model, at half-filling, 
with a canonical transformation, 
with which the charge and sSC pair correlations are mapped to the 
spin correlation in the repulsive side as
\begin{eqnarray}
U&\leftrightarrow&-U\nonumber\\
\langle (n_{i\uparrow}+n_{i\downarrow})(n_{j\uparrow}+n_{j\downarrow})\rangle
&\leftrightarrow&
\langle (n_{i\uparrow}-n_{i\downarrow})(n_{j\uparrow}-n_{j\downarrow})\rangle\nonumber\\
&&\propto
\langle S_i^zS_j^z\rangle\nonumber\\
\langle c_{i\downarrow}c_{i\uparrow}c_{j\uparrow}^\dag c_{j\downarrow}^\dag\rangle
&\leftrightarrow&
\langle c_{i\downarrow}c_{i\uparrow}^\dag c_{j\uparrow}^\dag c_{j\downarrow}\rangle\nonumber\\
&&\propto
\langle S_i^-S_j^+\rangle.
\label{eqn:Nagaoka}
\end{eqnarray}
This implies that the charge and sSC pair correlations are degenerate
and both decay like $r^{-1}$ as far as the Hubbard model
($U<0$) and the Hubbard-Holstein model ($\tilde U\equiv U-\lambda<0$)
in the anti-adiabatic limit are 
concerned.

If we go back to the Hubbard-Holstein model, 
the power-law fit for an exponentially
decaying correlation gives a finite exponent for finite systems, so
what we should observe when a correlation does become power-law
is a 
decrease of the exponent around the phase transition point.
We observe in Fig. \ref{fig:1dexp_5.0} that the exponent for the SDW
(CDW and sSC) decreases (increases) as $\tilde U \equiv U-\lambda$
is increased from the negative to positive sides.
This behavior is similar to the Hubbard model having 
$\tilde U$ as the on-site interaction, except that the exponent
for the charge correlation is slightly smaller than that
of the sSC pair correlation
in general for finite $\omega_0/t$
as mentioned above. This is
similar to the case of the Holstein model $U = 0$
with an $\omega_0/t\sim 5$ (not shown).

\subsection{Phonon numbers per site in the Hubbard-Holstein model}
\begin{figure}
\begin{center}
\includegraphics[width=8.6cm]{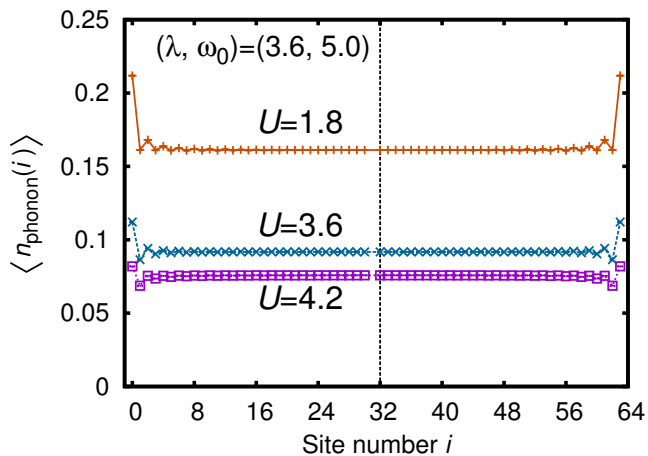}
\caption
{(Color online) Phonon number $\langle n_{\rm phonon}(i)\rangle$
plotted against $i$ for $U = 1.8, 3.6, 4.2$ and $(\lambda,\omega_0) = 
(3.6, 5.0)$ 
for a 64-site, half-filled Hubbard-Holstein chain.
}
\label{fig:1d_phnumbervsi}
\end{center}
\end{figure}

We can calculate the phonon occupation number for each site 
$n_{\rm phonon}(i)$ in the ground state $\psi_0$, defined as
\begin{equation}
\langle n_{\rm phonon}(i)\rangle
\equiv
\langle \psi_0|a_i^\dag a_i|\psi_0 \rangle.
\label{eqn:phnumber}
\end{equation}
We can check the boundary effects 
by plotting the phonon number against the site number $i$ in
Fig. \ref{fig:1d_phnumbervsi}.
The curves are almost flat except for a few sites around either edge 
of the chain.

\begin{figure}
\begin{center}
\includegraphics[width=8.6cm]{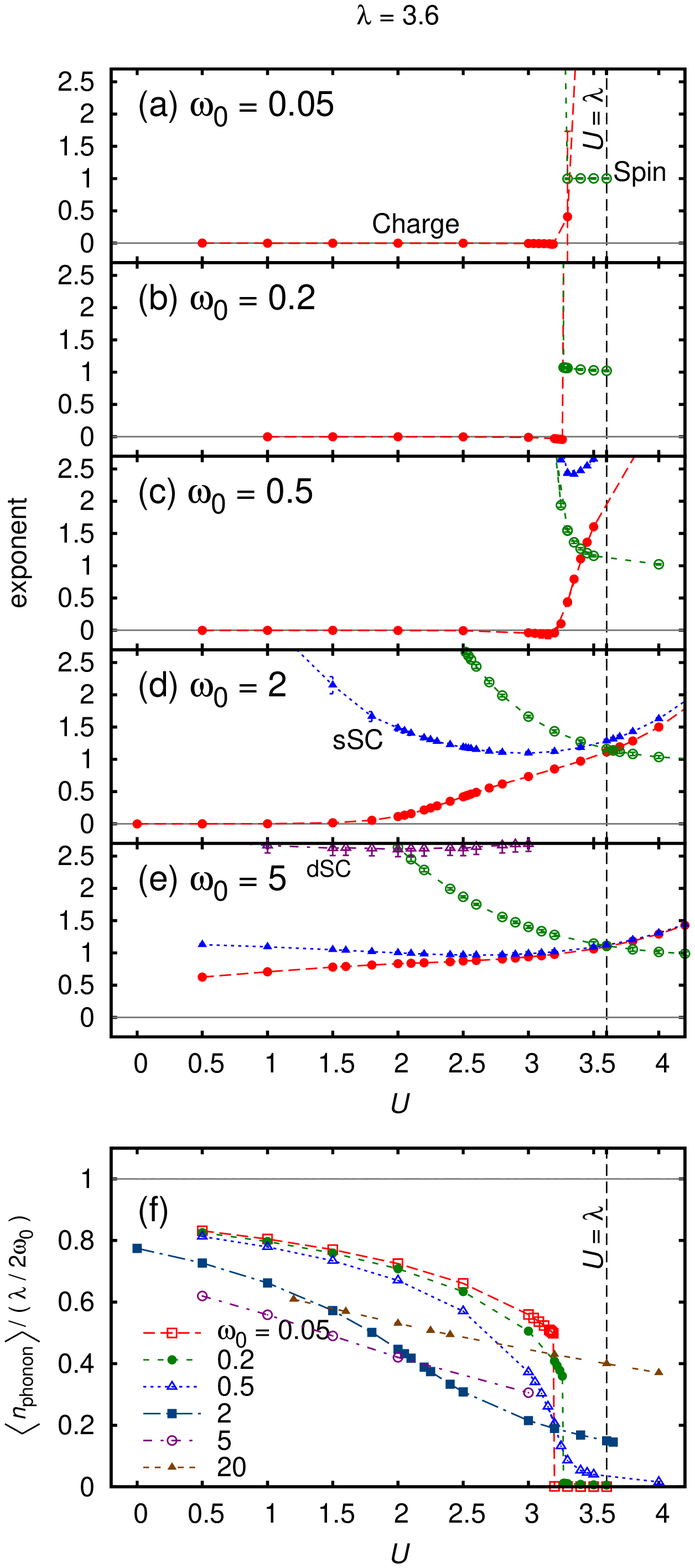}
\caption{(Color online) (a) -- (e) Exponents of correlation functions
determined by calculating exponents for a 20-site Hubbard-Holstein model
for $\lambda/t=3.6$ with various values of $\omega_0/t = 
0.05 - 5$.  
(f) The phonon number per site, normalized by eqn.(\ref{eqn:phononnumber}).}
\label{fig:exp3.6}
\end{center}
\end{figure}

We plot the phonon number at the center of the chain,
$\langle n_{\rm phonon}\rangle$, as a function of $U$
in Fig. \ref{fig:exp3.6}(f).
Phonons mediate an attraction 
between two electrons with opposite spins on the same site, which 
amounts to $\lambda\equiv2g^2/\omega_0$
in the $\omega_0\rightarrow\infty$ limit.
For a finite $\omega_0$,
the attraction $\lambda$ is smaller than this asymptote $\lambda$.\cite{2005PhRvL..94b6401S,2005PhyB..359..636C}
For $U\ll\lambda$, two electrons tend to form a bipolaron,
where the electrons are strongly bound to each other by sharing phonons they excite.
Bipolarons, at half-filling, 
tend to occupy every second site to reduce kinetic energy.
In this case the charge correlation that corresponds to the $2k_F$ CDW
is the strongest.
The number of excited phonons per site increases as $U$ is decreased,
because the fluctuation from the single occupancy increases as
the CDW correlation becomes stronger.

The limit of immobile bipolarons can be understood as follows:
we consider two electrons on a site in (\ref{eqn:HHdiff}) and neglect their hopping,
to apply a Lang-Firsov type transformation $\hat{a}=a-\sqrt{\alpha}$.
Then the local Hamiltonian for the phonon on this site is
\begin{equation}
H=g(a+a^\dag)+\omega_0 a^\dag a
=\omega_0\hat{a}^\dag \hat{a} + (\omega_0\alpha+2g\sqrt\alpha),
\end{equation}
with $\alpha\equiv (g/\omega_0)^2 = \lambda/2\omega_0$.
The second term is a constant, and
if we denote the $n$-phonon state as $|n\rangle$
($n=0,1,\ldots$),
the ground state $|\psi_0\rangle=\sum_n b_n|n\rangle$ satisfies
\begin{equation}
\hat{a}|\psi_0\rangle = (a-\sqrt\alpha)\sum_n b_n|n\rangle
= \sum_{n\geq1}(\sqrt n b_n-\sqrt\alpha)|n-1\rangle = 0.
\end{equation}
This is solved as a coherent state,
\begin{equation}
b_n=\sqrt{e^{-\alpha}\alpha^n/n!},
\end{equation}
which gives
\begin{eqnarray}
\lim_{U\rightarrow 0}\langle n_{\rm phonon}\rangle
&=& \langle a^\dag a \rangle
= \frac{\sum_{n\geq1}\alpha^n/(n-1)!}{\sum_{n\geq0}\alpha^n/n!}
\nonumber\\
&=& \alpha = \lambda/2\omega_0.
\label{eqn:phononnumber}
\end{eqnarray}
We can observe in Fig. \ref{fig:exp3.6}(f)
that the system indeed approaches to this limit as $U$ is decreased.

A deviation observed for larger values of $\omega_0$
in Fig. \ref{fig:exp3.6}(f) is interesting 
in that the region is intermediate between
the $U\ll\lambda$ and $U\gg \lambda$ limits.
There we can expect a behavior different from the limiting behavior.

\subsection{The case of smaller $\omega_0/t$}
\begin{figure}
\begin{center}
\includegraphics[width=8.6cm]{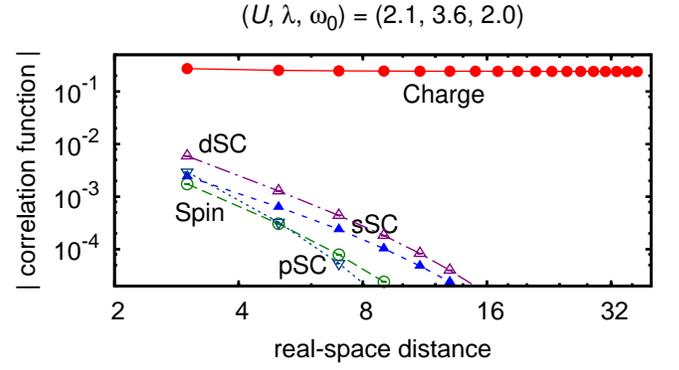}
\caption
{(Color online) Correlation functions for $(U, \lambda, \omega) = (2.1, 3.6, 2.0)$
on a 64-site, half-filled Hubbard-Holstein chain with $n_B=6$,
where CDW correlation is dominant and almost constant.}
\label{fig:1dcor_CDWconst}
\end{center}
\end{figure}

When the value of $\omega_0/t$ is smaller than about 3,
we start to observe in Fig. \ref{fig:exp3.6}(c)
behaviors of the correlation functions different from those discussed above: 
the value of $U$ at which the exponent for the charge and spin correlations
coincide becomes considerably smaller than $\lambda$. 
So \textit{the $(U-\lambda)$ Hubbard model picture is no longer valid}.
Secondly, the charge correlation
becomes almost flat against distance as in Fig. \ref{fig:1dcor_CDWconst}, 
implying a charge-ordered (CO) phase, in some region of $U$ 
in Fig. \ref{fig:exp3.6}(a)--(d),
while this does not happen for the on-site pair correlation.

To analyze the nature of this behavior, 
we can again look at the phonon occupation number against $U$ 
for various values of $\omega_0$ for a fixed value of $\lambda$
(Fig. \ref{fig:exp3.6}(f)).
For small enough $\omega_0$ the phonon occupation number 
abruptly increases at some value $U_c$ as $U$ is decreased.  
We can see that 
the exponential decay of all the correlations except the charge 
correlation in fact start around $U=U_c$.  
Thus we identify that this is where the SDW -- CDW transition occurs.
For larger $\omega_0$, on the other hand, the change in both
the exponents and the phonon occupation number are gradual. 
The intermediate values 
($0 < \langle n_{\rm phonon}\rangle < \lambda/2\omega_0$) 
of the phonon occupation number
is consistent with the power-law behavior of the CDW and sSC
exponents, which corresponds to a metallic phase with a closed charge gap,
where the fluctuation in the local electron number 
(hence the fluctuation in the local phonon number) will be large.

\begin{figure}
\begin{center}
\includegraphics[width=8.6cm]{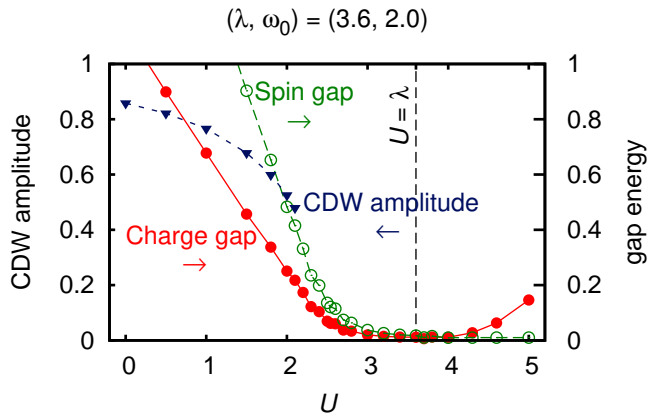}
\caption
{(Color online) Charge gap, spin gap, and CDW amplitude plotted against $U$
for the half-filled Hubbard-Holstein model
with $(\lambda, \omega_0) = (3.6, 2.0)$.
Data obtained for $L = 20, 30, 40, 64$ have been linearly extrapolated
to $1/L\rightarrow0$.
}
\label{fig:CDW_gap}
\end{center}
\end{figure}

When we look at the CDW correlation,
\begin{equation}
\xi_{\rm Charge}(i,j)\equiv \left\langle\psi_0\left|\left(\sum_\sigma n_{i\sigma}\right)
\left(\sum_\sigma n_{j\sigma}\right)\right|\psi_0\right\rangle-1,
\end{equation}
its amplitude turns out to either decay, or remain 
almost constant ($\xi_{\rm CO}$) against the distance $r=|i-j|$.  
Since $\langle \sum_{\sigma}n_{i\sigma}\rangle$ alternates 
between $1+f_{\rm CO}$ and $1-f_{\rm CO}$ in the latter case 
at half-filling, we can simply define the charge-ordering amplitude 
$f_{\rm CO}=\sqrt{|\xi_{\rm CO}|}$.
We plot $\Delta_{\rm charge},\Delta_{\rm spin}$
and $f_{\rm CO}$ against $U$ in Fig. \ref{fig:CDW_gap}.
As $U$ is decreased in the CO region, we observe 
that (i) the CDW amplitude $f_{\rm CO}$ increases and approaches unity,
(ii) both the spin and charge gaps increase almost linearly,
and (iii) the spin gap is nearly twice the charge gap.
For the CDW/sSC region, the charge gap almost vanishes.
The spin gap becomes also small, but remains larger than the charge gap. 
Finally, when $U$ exceeds $\lambda$, the spin gap remains zero, 
while the charge gap slightly grows with $U$.

The effect of the ratio $\omega_0/t$
on the $U$-dependence of $\langle n_{\rm phonon}\rangle$ can now
be understood in the limits of $\omega_0\gg t$ and $\omega_0\ll t$ as follows:
\begin{enumerate}
\item In the anti-adiabatic limit ($\omega_0\gg t$), the response of the
phonons to the motion of electrons is so fast that
we can think of the role of phonons as just mediating the attraction, $\lambda$,
between electrons.
Electrons then tend to occupy different sites for $U>\lambda$,
with spins antiparallel between neighbors because of the exchange interaction.
In this case the spin correlation that corresponds to the $2k_F$ SDW
is the strongest, and the number of excited phonons per site 
decreases with $U$, 
because the electron occupancy becomes closer to unity for every site.

\item In the adiabatic limit ($\omega_0\ll t$), 
for $U\ll \lambda$ the deviation ($f_{\rm CO}$) 
of the electron occupancy from the average value is large, so that 
a larger number of phonons are excited.  
The dynamics of phonon is much slower than that of electrons 
for $\omega_0\ll t$, and exactly when the charge-ordered states exist 
depends on $\omega_0$ as we shall show on the phase diagram.  
For $U\gg \lambda$ electrons cannot move and form a Mott insulator, 
so that phonons are hardly excited.  
In the region $\omega_0 \sim < t$ we observe a jump in $\langle n_{\rm phonon}\rangle$
as a function of $U$.
This should correspond to the CDW transition.
\end{enumerate}

\subsection{The phase diagram}
As we have seen above,
we have two to three distinguishable regions
out of the five phases compared in Table \ref{tbl:phases}
when we increase the value of $U$ from $U=0$:
\begin{table}
\begin{center}
\caption{Summary of possible phases at half-filling.
We denote power-law and exponential decays of
correlation functions (`corr.') as `power' and `exp.', respectively.
}
\label{tbl:phases}
\renewcommand{\arraystretch}{1.5}
\begin{tabular}{lcccc}
\hline\hline
phase & CO & CDW/sSC & SDW\\\hline
spin corr. & exp. & exp. & power\\
charge corr. & long-range order & power (smallest $\eta$) & exp. \\
sSC corr. & exp. & power & exp.  \\
charge gap & finite & 0 & finite  \\
spin gap & finite & finite & 0 \\
$\langle n_{\rm phonon}\rangle$ & $\rightarrow \lambda/2\omega_0$ & intermediate & $\rightarrow 0$\\\hline\hline
\end{tabular}
\end{center}
\end{table}
\begin{enumerate}
\item Charge ordering (CO) ($\Delta_{\rm spin}\sim\Delta_{\rm charge}>0$) ---
The charge correlation is dominant and does not exhibit
a significant decay against distance.
Other correlation functions decay exponentially.
The charge and spin gaps are still large,
but decrease with $U$.
These should be the characteristics of a long-range charge order,
which is not forbidden from \cite{PhysRevLett.17.1133, PhysRev.158.383}.
We denote this region as CO.
\item CDW/sSC ($\Delta_{\rm charge}\ll t, 0<\Delta_{\rm spin}\ll t$) ---
The charge and sSC correlations decay with power-law, with
the charge correlation the stronger of the two.
The spin gap is finite (although a size-scaling argument will be required
to quantify this),
while the charge gap is closed (although a size-scaling argument will again
be required).  
We denote this region as CDW/sSC because the sSC
correlation, while not dominant anywhere, decays with power-law as well.
The dSC correlation
(nearest-neighbor spin-singlet pair) is even less dominant.
\item SDW ($\Delta_{\rm charge}>0, \Delta_{\rm spin}\ll t$) ---
Only the spin correlation decays with power-law with
the exponent being nearly unity.
The spin gap is closed or much smaller than $t$.
The charge gap becomes larger as $U$ is increased.
We denote this region as SDW.
\end{enumerate}
\begin{figure}
\includegraphics[width=8.6cm]{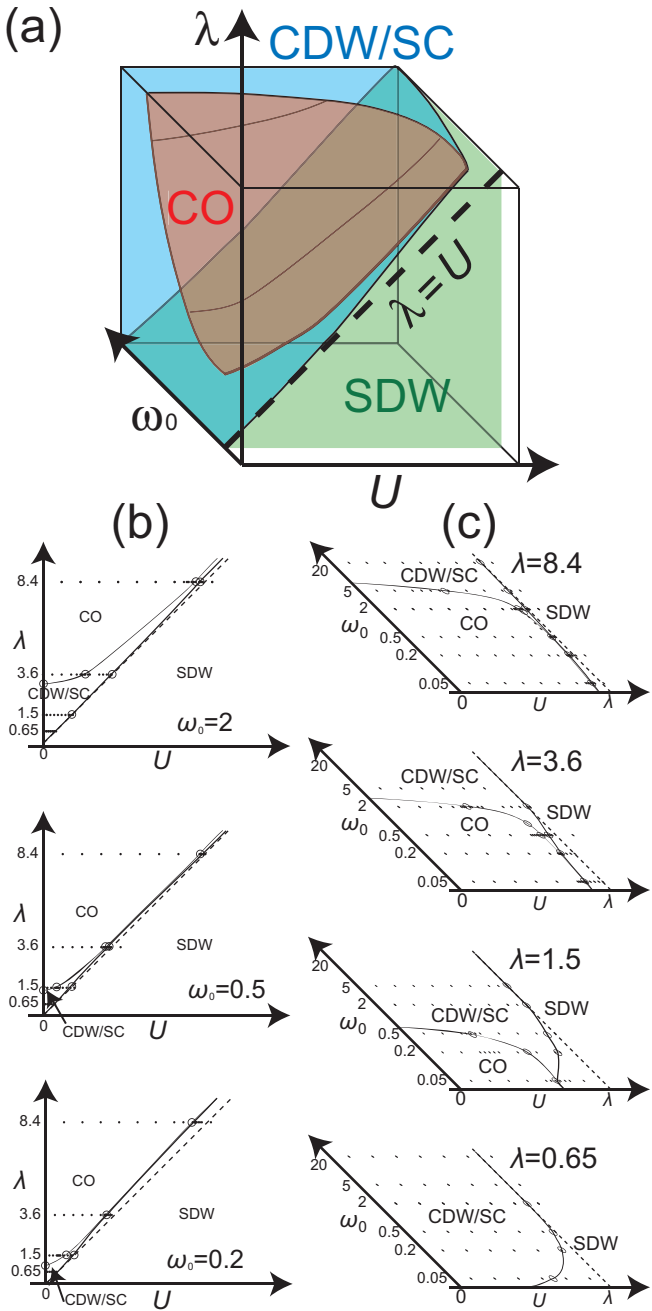}
\caption{(Color online)
(a) The summary phase diagram on the whole parameter space for the half-filled Hubbard-Holstein model as derived from
(b) the results obtained for the $U-\lambda$ cross sections, and (c) $U-\log \omega_0$ cross sections.
The boundaries have been determined from the data points marked with dots.}
\label{fig:UomegaPhase}
\end{figure}

We draw phase diagrams 
on the $U-\lambda$ plane for various values of
$\omega_0 =0.2, 0.5, 2$, and
on the $U-\omega_0$ plane
for various values of $\lambda = 0.65, 1.5, 3.6, 8.4$
(Fig. \ref{fig:UomegaPhase}(b)(c)).  
We observe on the $U-\omega_0$ plane that
the CDW/SC region is narrow for $\omega_0 \ll \lambda$,
while the region expands with the CO-CDW/SC boundary shifting
to smaller $U$ and the CDW/SC-SDW boundary shifting to larger $U$,
until the region finally extends up to $U\sim \lambda$ when
$\omega_0 > \sim \lambda$.
The CO region also becomes wider for smaller $\omega_0$ or
smaller $U$, but vanishes at $\omega_0\sim \lambda$.
For $\lambda=0.05$ we have not detected this region.

On the $U-\lambda$ plane,
we observe that the SDW region covers all of $U>\lambda$,
but it extends beyond the line $U=\lambda$ when $\omega_0$ is smaller.
The CDW/sSC region is broader for larger values of $\omega_0$,
and we have not detected CO region for $\omega_0/t=20$.  
Figure \ref{fig:UomegaPhase}(a) summarizes the whole region.

\section{Doped chain}
\label{sec:doped}
\begin{figure}
\begin{center}
\includegraphics[width=8.6cm]{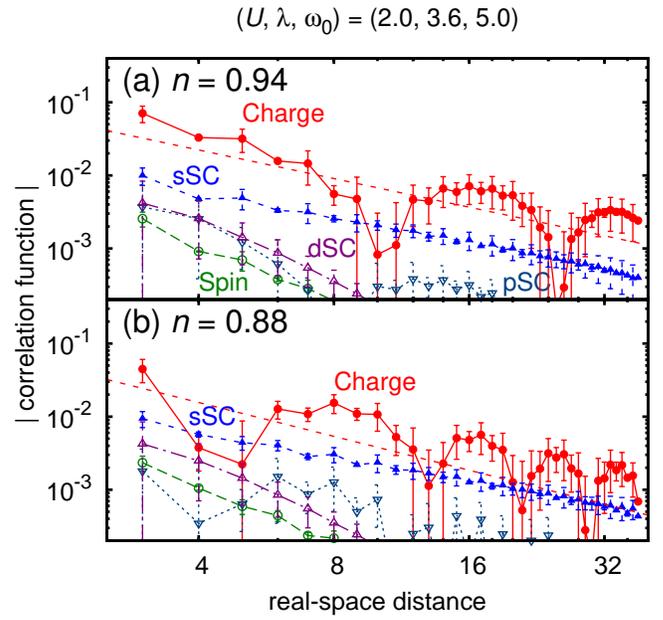}
\caption{(Color online) Correlation functions for a 64-site, {\it hole-doped} 
[$n=30/64=0.47$ in (a) and $n=28/64=0.44$ in (b)] Hubbard-Holstein chain
with $(U, \lambda, \omega_0) = (2.0, 3.6, 5.0)$.
The exponents for the sSC correlation (dotted straight line)
are estimated to be
$\eta_{\rm sSC}=1.16\pm0.02$ in (a) and $\eta_{\rm sSC}=1.09\pm0.02$ in (b),
while the CDW correlation to be
$\eta_{\rm CDW}=1.30\pm0.10$ in (a) and $\eta_{\rm CDW}=1.54\pm0.13$ in (b).
}
\label{fig:DopedExp}
\end{center}
\end{figure}
Having seen that the pairing correlation is only subdominant
in the half-filled Hubbard-Holstein model, we can naturally
ask the question:
can the system away from the half-filling be superconducting?
Since the effect of doping is dramatic
in purely electronic systems such as the Hubbard model,
this is an obviously interesting avenue to examine.
For the doped systems we change in our DMRG calculation 
the target quantum number,
which is the pair of the up spin and the down spin electrons
$(N_\uparrow,N_\downarrow)$, 
and change the values of $\overline{n}_{\sigma}$ in (\ref{eqn:HHdiff})
accordingly.

The functional forms of the correlations for the non-half-filled
system have turn out to have the following
property: the pair correlation functions decay almost
exactly with a power-law, while the charge correlation decays as a power-law
multiplied by an oscillating factor.
The dominant wavenumber for the CDW correlation should be 
$2k_F$, which equals to $\pi/a$ ($a$: the lattice constant) at half-filling, 
but becomes larger as the hole doping is increased.
In fact, we can fit the result for the CDW correlation as
\begin{equation}
\xi_{\rm CDW}\simeq C r^{-\eta_{\rm CDW}} \cos\left(q r + c\right),
\end{equation}
where $\eta_{\rm CDW}$ is the exponent, $C$, $c$ and $q(\simeq n\pi/a)$
are fitting parameters.

We compare the CDW exponent thus obtained with the exponent 
for the sSC correlation (dashed line) in Fig. \ref{fig:DopedExp}.
The calculated exponents for CDW and sSC are similar in Fig. \ref{fig:DopedExp} (a),
where the electron filling $n$ is close to unity.
However, sSC becomes clearly dominant over CDW for a larger level 
of hole doping ($n=0.44$) in Fig. \ref{fig:DopedExp} (b).
The exponents in this case are: $\eta_{\rm sSC}=1.09\pm0.02$ and
$\eta_{\rm CDW}=1.54\pm0.13$.  
So this is the key result for the doped system:
the sSC correlation has a smaller exponent, and thus
dominant in the sufficiently doped case.

The doped system further poses an interesting problem:
for purely electronic systems such as the Hubbard model, 
doping the Mott insulator in two or higher dimensions
favors superconductivity.
For the Hubbard-Holstein model with
phonons, however, the situation should be far from trivial, especially
because the simplified 
$\tilde{U} \equiv U - \lambda$ picture can be invalidated 
as stressed above.  So this should be a further avenue for 
future studies.

\section{Effect of the lattice structure --- Trestle lattice}
\begin{figure}
\begin{center}
\includegraphics[width=3.6cm]{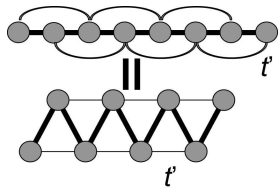}
\caption{A trestle lattice (bottom) is equivalent to a chain
that has the next-nearest neighbor hopping $t'$ (top).}
\label{fig:trestlelattice}
\end{center}
\end{figure}
A second approach to make the pairing correlation dominant
should be, in our view, to modify the electronic band structure.
Along the line of argument above, we can specifically 
break the electron-hole symmetry to lift the 
degeneracy between CDW and sSC correlations, eqn.(\ref{eqn:Nagaoka}), 
which exists in the Hubbard model \cite{1974PThPh..52.1716N} 
and is shown above to persist for the Hubbard-Holstein model
in the anti-adiabatic limit. 
This can be done by considering quasi-1D lattices
such as ladder or trestle lattices.  A trestle lattice, for
instance, is equivalent to a chain where a next-nearest neighbor hopping $t'$
is introduced (Fig. \ref{fig:trestlelattice}).
This washes out the bipartite symmetry in the
band structure, which in turn breaks the electron-hole symmetry.
We can then seek the possibility of making the pairing correlation dominant, 
even at half-filling, in the Hubbard-Holstein model,
as have been done for the Hubbard model.\cite{1996PhRvB..5410054F,1997JPSJ...66.3371K}

\begin{figure}
\begin{center}
\includegraphics[width=8.6cm]{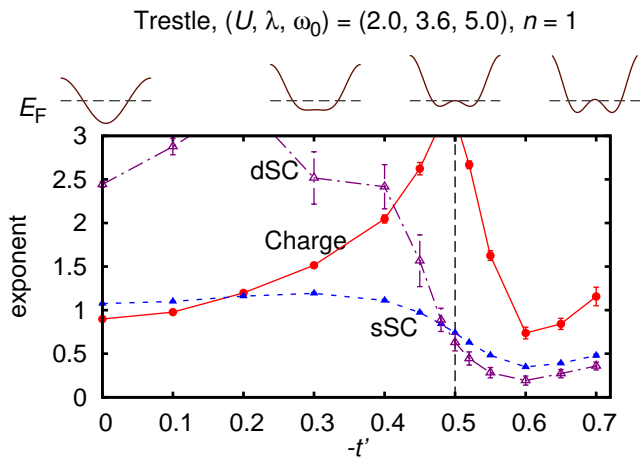}
\caption
{(Color online) Correlation exponents
plotted against $-t'$ for a 40-site, half-filled Hubbard-Holstein model
on the trestle lattice with $(U, \lambda, \omega_0) = (2.0, 3.6, 5.0)$.
At least $m=630$ states have been retained in the last (10th) sweep
of the finite algorithm DMRG. 
The vertical dashed line represents the boundary at
which the number of Fermi points changes from two to four 
as indicated by the band dispersion (top insets).
}
\label{fig:exptrestle}
\end{center}
\end{figure}

So we have performed the DMRG study for the trestle lattice.
We retained up to $m=720$ states per block. The maximum truncation
error in the final sweep is around $10^{-5}$, so it is considerably
larger than in the case of the simple chain.
\begin{figure}
\begin{center}
\includegraphics[width=8.6cm]{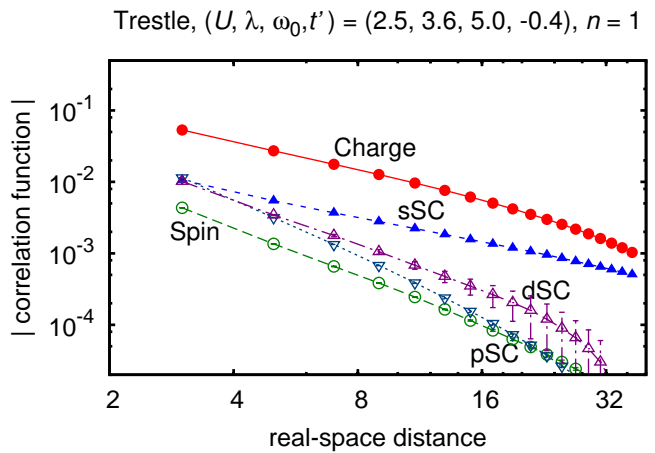}
\caption{(Color online) Correlation functions
plotted against the real-space distance $r$
for a 64-site, half-filled Hubbard-Holstein model on the trestle lattice
with $(U, \lambda, \omega_0, t') = (2.5, 3.6, 5.0, -0.4)$.}
\label{fig:corrtrestle}
\end{center}
\end{figure}
If we look at the correlation functions 
against the real-space distance in Fig. \ref{fig:corrtrestle}, 
we can see that the sSC correlation is indeed
dominant at $(U, \lambda ,\omega_0, t') =  (2.5, 3.6, 5.0, -0.4)$,
even though we consider a half-filled system.
Here both of the charge and sSC correlation functions exhibit
power-law decay times oscillating factors.
\begin{figure}
\begin{center}
\includegraphics[width=8.6cm]{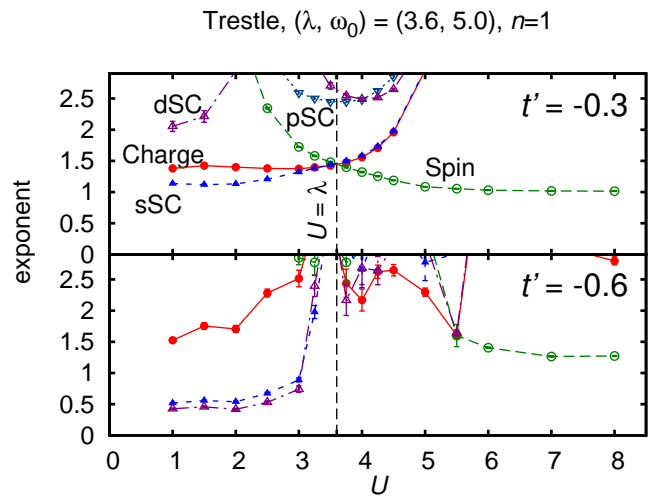}
\caption{(Color online) Exponents of correlation functions
plotted against $U$ for a 64-site, half-filled Hubbard-Holstein model
on the trestle lattice for $(\lambda, \omega_0) = (3.6, 5.0)$ with 
$t'/t = -0.3$ (top panel) or $-0.6$ (bottom).}
\label{fig:expUtrestle}
\end{center}
\end{figure}
We can then
look at the exponents of correlation functions against $U$ for various values
of $t'/t$, which controls the band structure 
(top insets of Fig.\ref{fig:exptrestle}).  
A special interest is that the number of Fermi points at half-filling for the 
non-interacting system increases from two to four at $|t'|>0.5$.  
Curiously enough, the case of four Fermi points is seen in the bottom 
panel of Fig. \ref{fig:expUtrestle} to have a 
dominant dSC correlation for $U\ll \lambda$.  
This is the key results for the trestle lattice.

For $U\sim \lambda$ we observe relatively larger error bars in the exponents.
This should be due to the presence of four Fermi points, for
which we have a larger number of $k$-points around $E_F$ in a finite system.
When $U\gg \lambda$, dominance of the SDW is recovered
as in the case of the 1D chain and trestle lattice with smaller $|t'|$.

Next we plot the exponents of the correlation functions
as functions of $t'$ in Fig. \ref{fig:exptrestle}.
The vertical dashed line in the figure represents the boundary at
which the number of Fermi points changes from two to four.
In other words, a geometrical frustration (i.e., interference between
the nearest-neighbor and second-neighbor transfers) becomes the
strongest around this boundary.
We can notice that the pairing correlations (sSC and dSC) 
tend to be dominant around the boundary, while 
the CDW correlation begins to decay faster there.

\section{Summary}
To summarize we have shown for the 1D Hubbard-Holstein model the following: 
(i) For the half-filled case we have obtained the phase diagram
for the whole parameter space spanned by the Hubbard $U$,
phonon frequency $\omega_0$, and the electron-phonon
coupling $\lambda$.
A region is shown to exist between the SDW and CO
phases, where the superconducting correlation is only subdominant against
CDW.
(ii) When we either (a) dope the electronic band, or (b) change
the electronic band structure by considering a trestle lattice,
the on-site pair correlation,
and in the case of (b) the nearest-site singlet pair correlation,
indeed become dominant over CDW.
This is to be contrasted with
the Tomonaga-Luttinger picture ($g$-ology), 
which would dictate that the pair correlation
should not dominate when there are two Fermi points
(the region left of the dashed line in Fig. \ref{fig:exptrestle}).  

\section*{Acknowledgment}
We wish to thank Yasutami Takada for illuminating discussions.
M.T. thanks Eric Jeckelmann, Eugene Demler and Giorgio Sangiovanni
for helpful comments.
This work is in part supported by a Grant-in-Aid for Science Research on
Priority Area ``Anomalous quantum materials''
from the Japanese Ministry of Education.
\bibliography{full3}
\end{document}